\documentclass[aps,multicol,psfig,epsf,epsfig,preprint,showpacs]{revtex4}
\usepackage{graphicx}
\usepackage{amssymb}
\usepackage{amsmath}
 
\begin{document}
 
\title{Crackling noise in paper peeling} 

\author{L.I. Salminen$^1$, J.M. Pulakka$^1$, J. Rosti$^1$, 
M.J. Alava$^{1,2}$, and K.J. Niskanen$^3$} 

\affiliation{$^1$Helsinki University of Technology, Laboratory of Physics,\\
P.O.Box 1100, FIN-02015 HUT, Finland }

\affiliation{$^2$SMC-INFM, Dipartimento di Fisica,
Universit\`a "La Sapienza", P.le A. Moro 2
00185 Roma, Italy}

\affiliation{$^3$Finnish Pulp and Paper Research Institute, P.O.Box 51, 
FIN-02151 Espoo, Finland}

\begin{abstract}
Acoustic emission or crackling noise is measured from an experiment
on splitting or peeling of paper. The energy of the events
follows a power-law,
with an exponent $\beta \sim 1.8\pm 0.2$. The event intervals
have a wide range, but superposed on scale-free statistics
there is a time-scale, related to the typical
spatial scale of the microstructure (a bond between two
fibers). Since the peeling takes place via
steady-state crack propagation, correlations can be
studied  with ease and shown to exist in the series of acoustic events. 
\end{abstract}
\maketitle
\date{\today}

There are several experimental signatures of power-law statistics,
or scale-invariance, in fracture. One interesting case
is acoustic emission (AE),
which is produced by the release of small quantities of elastic
energy during the failure of an inhomogeneous sample, and is an
example of the phenomenon of ``crackling noise'', met in many
kinds of systems in physics \cite{sethrev,Bourev}. This
can result in analogies of the
Gutenberg-Richter's and Omori's laws for earthquakes 
\cite{Ciliberto,penn,Petri2,Lockner,oma}.
These are experimental observations; the former relates the probability
(for earthquakes or AE) of an event $P(E)$
to the energy or magnitude of the event $E$ by an exponent $\beta$. 
Correspondingly, the latter binds by an exponent $\alpha$ the waiting time
$\tau$ between the events and their probability $P(\tau)$.  Very often
these laws are witnessed simultaneously in both microfracturing
and earthquakes \cite{simu}. The origins of both the
power-laws are mostly unknown and the same holds also for
the possible connection between these.

The common idea about material failure is based on a critical
defect:  the sample fails catastrophically once the local strength, 
that follows from among others the size of the flaw, is exceeded.
The observation of AE implies  often that microscopic damage 
is being created, contrary to one, single catasthropic
crack growth event. One consequence is that, in analogy
to earthquakes again, the idea of predictablity becomes
of interest. This is due to the precursors to final
failure that could be diagnosed by AE. In general terms
it is of fundamental materials science and statistical
physics interest that universal or critical fracture behavior should 
exist in the presence of varying material properties, like anisotropy,
related to the shape of the stress-field close to microcracks or
a notch, or to e.g. asymmetric disorder \cite{lopinaa}. 
Any power-laws in AE or crackling noise should originate
from a basic, fundamental mechanism, in spite of such complications.

One particulary inviting proposal
is to follow the passage of a crack front through a quasi 
two-dimensional sample, which can be realized eg. in a
weak interface between two three-dimensional elastic plates \cite{line,Fisher}.
This scenario has the advantage, over say ordinary tensile
tests, that the crack propagation takes
place in a {\em steady-state} in contrast to most other experiments
of fracture. In plexiglass, the front may be self-affine,
and a roughness exponent $\chi$ close to 0.6 has been measured,
which is intriguing given that the advancement takes
place in avalanche-like events \cite{maloy}. The theoretical understanding
of the phenomenon is based on numerical models and on stochastic
equations for the crackline dynamics. These have so far not
been able to account for most of the observed features, nor
is there any understanding of whether the roughening could
be expected to be ``universal''.

\begin{figure}
\includegraphics[width=7cm
]{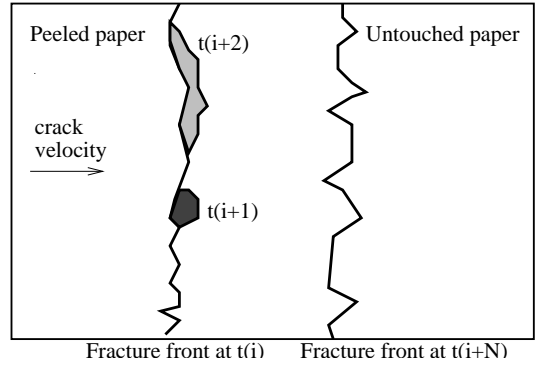}
\caption{Fracture propagation schematics in the peel-in-nip test.} 
\label{figkaksib}
\end{figure}

In this work we look at the statistics of fracture AE
in a set-up that mimicks such line dynamics in ordinary paper
by peeling a sheet into two, as a crack advances through it
(Figs. \ref{figyksi} and \ref{figkaksib}). 
We obtain that the energy of AE events scales as a power-law, 
like in tensile tests, which produce analogies
of both the Gutenberg-Richter and Omori's laws.
This takes place with
a novel exponent which is much larger than those usually
met in fracture AE. The temporal statistics exhibit 
a wide spectrum of intervals between the events. Translated
into the average distance that the fracture line propagates
we show below that there is a typical scale, related to 
the size of a fiber-to-fiber bond. 
Rupture processes often involve time-dependence, but at least
for low strain rates the structural scale stays the same.
The dynamics of the peeling line is most likely
quasi-one-dimensional. That is, due to the thinness of a
sheet of a paper the stress field is not able to penetrate
into the intact paper so as to give rise to two ``scalefree''
lengthscales \cite{zapperi}.  We also compare to other kind of AE data
for the same materials, demonstrating that for localized
fracture, the $\beta$-exponent is much larger, and that
in a tensile test values met before can be obtained \cite{oma}. 
Such experimental data  should be useful to  
to formulate and test microscopic fracture models.

In the case of paper, Yamauchi made pioneering acoustic emission 
measurements \cite{yama}, with the claim that fiber and bond
breakages can be distinguished in the amplitude histogram
of the signal. 
The idea of basic scales
should be contrasted with Fig.~\ref{figkaksib} that introduces
the crack advancement scenario. Due to the nature
of the set-up,  the stress-field
is expected to be cut-off quickly (in some models exponentially)
with increasing distance from the average crackline position.
Thus the individual
events, interpreted in terms of the area over which the ``avalanche''
passes, are in practice one-dimensional. Of course, the crack can
fluctuate on very small scales in the $z$-direction, perpendicular
to the sheet.

With intact samples in a tensile test the strength
depends on weak spot statistics. In practice some existing weak region
often launches the crack growth, and after that rupture becomes mainly
a local problem.  Naturally, if a sizeable initial (edge or center)
notch is applied, the fracture process focuses in the 
so-called fracture process zone, FPZ,
around the crack-tip (for paper, the characteristic dimension of 
FPZ is upto 5 mm, typically).
Since fracture can happen in a diffuse or in such
a localized way one might see two separate statistics related to
these processes, and distinguish two regimes. These would be the
pre-fracture phase, when behavior is nearly elastic and cracking or
damage disperse and the second part taking place after and at the
stress maximum. During this regime a single crack is propagating and 
the failures concentrate in the FPZ, the vicinity of the crack-tip.

In the presence of disorder the fracture can be an irregular process of
elementary rupture events separated by interarrival or waiting times,
and spread out geometrically in the sample
The variety of theoretical or computer models available
differ in the level and details of load resharing after microfailures
\cite{Zapperi}. 
The events in such studies consist of  
single spring-like breakages, with a varying number of those elementary
breakages within one coherent event. In the 
case the fracture takes place via the avalanche-like dynamics,
2d computer simulations indicate a $\beta$-value of 1.7 \cite{recent}.
Also, $\alpha = 0.94 \pm 0.20$ was reported in a simulation of hydraulic
fracture \cite{herrmann}. All such studies do not apply to the peel-experiment
at hand, ie. there are no theoretical results pertaining to the
possible AE statistics in our case. 

\begin{figure}\includegraphics[width=7cm
]{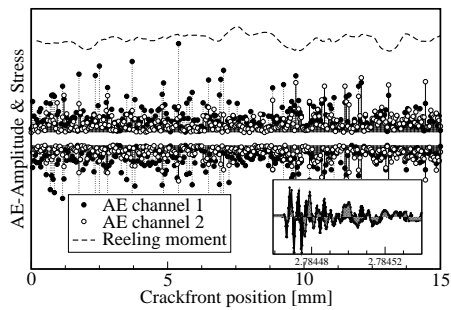}
\caption{Example of the data in peel-in-nip test. 
Both the AE channels, and the force or reeling moment
signals are included. The former is in arbitrary units,
for the latter see text.} \label{figkaksic}
\end{figure}

Acoustic emission as such is a well-known technique to monitor fracture in
e.g. composites.  The rapid release of elastic energy can be observed by
ultrasonic sensors \cite{corte}, with little influence on the actual
fracture process. 
The peel-in-nip method to split paper is based on the 
nip between two rolls rotating synchronously (Fig. \ref{figyksi}).
The front end of the sample is attached to both the
rolls by tape and then the cylinders are reeled to initiate the peeling.
Then the cleavage proceeds based on an equilibrium of three supporting
forces. The peeling takes place near the nip so that angle between the
planes of the halves is close to $\pi$ and the actual value
has a slight dependence on the particular kind of paper.
The test produces very large fracture surfaces; over 100 mm${}^2$ compared to
a typical area 1 mm${}^2$ in standard tensile tests.
Paper is a fibrous material, such that the fibers
form interpenetrating layers.
In standard tests the fibers are loaded mostly in-plane,
and several microscopic failure mechanisms co-exist (bond
or fiber breakages, fiber pull-out). 
The fractureline separating the intact part and the separated
halves is continuous, however the paper structure is discrete on
scales below the fiber length. Thus some bond dimension related
cross-over in any distribution (energy or intervals of events, or durations)
is possible.  Figure \ref{figkaksic} demonstrates a typical
test: note both that there is a maximum scale for interarrival
times, and that, of course, the AE event size has no trend during the test.

We used two sets of handsheets, paper made in a standard laboratory
mould from standard mechanical pulp. Typical tensile strength is
3 $kN/m$ and strain at break $2 \%$. 
One control parameter used is refining, or beating the pulp
which makes the fibers more flexible and more fibrillated.
Due to that the resulting paper is more uniform and becomes thus stronger.
Contrary to industrial paper the fiber orientation is uniform.
We had a set of handsheets with six different refining levels. 
Small (length 70 mm, width 15 mm) sample strips were used to reduce
the elastic energy often leading on catastrophic crack growth,
in ordinary tensile laboratory tests performed for comparisons.


During the experiment we acquire bi-polar acoustic amplitudes
simultaneously on two channels by piezocrystal sensors
(Physical Acoustic Corporation R15 tranducers, resonant
frequency 150 kHz)
as a function of time. In addition the force 
of peel-in-nip or tensile test was measured, typical values being
a few tens of Newtons depending on the thickness and rolling
friction of the grade of paper tested (in Fig. \ref{figkaksic}
about 90 N, with a variability of about 10 N) .
In tensile tests the two transducers were attached directly to paper and
no coupling agent was used.  Each channel has 12-bit resolution and
a sampling rate of 400 000/s. The transmission time from event origin to
sensors is order of 5 $\mu$s. The acoustic channels were first
fed through custom-made amplifiers,
and after that held using sample-and-hold circuit. The
shape of AE pulses can change and attenuate during transmission,
but  the effect should have has only minor effect to our analysis.
The acoustic time-series are reformed
offline by thresholding, by the detection of continuous and coherent events,
and by the calculation of event energy $E$, the sum of
squared amplitudes within the event. Events are separated by silent
(i.e. amplitude below threshold level) waiting intervals $\tau$. We do
not interpret the waveforms of the events but analyse the data in
the statistical sense.  The energy span is estimated to be
about 1 $\mu$J to 1 mJ; recall that the energy released
ends up as heat eventually.
In general the energy of the event is expected to be
proportional to the damaged area corresponding to the event
\cite{lys} and to the stress in that area. 

\begin{figure}\includegraphics[width=7cm
]{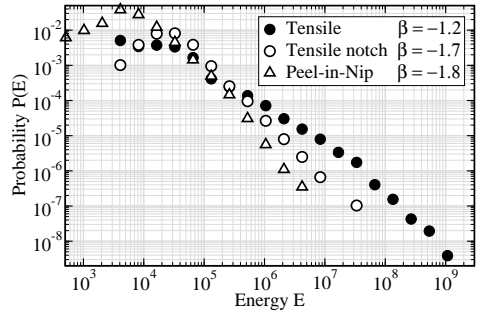}
\caption{Event energy distributions for the peel fracture
and comparisons.} \label{figkolme}
\end{figure}

The energy statistics shows that
acoustic events obey a Gutenberg-Richter -like
power-law for the peel-in-nip test (figure \ref{figkolme})
with the value $\beta=1.8 \pm 0.2$.
The figure provides as a comparison the energy data from
an ordinary tensile test:  the exponent $\beta=1.2\pm0.1$, in 
reasonable agreement with earlier such experimental values.
These can also be compared to the value
from a tensile test with a {\em large} initial notch,
resulting in the exponent $\beta = 1.7\pm0.2$. Note that
the peel-in-nip case presents some slight curvature.
The difference in the $\beta$-exponent between this
data and the tensile-with-notch appears us to be 
statistically reliable, in spite of the error bars.
Notice that as is usual in an AE test the energy scale
can not be calibrated quantitatively.

The different $\beta$ -values imply that there are scale-free
behaviors in both tensile and peeling fracture, in the same
material. Distributed damage all over the sample and microcrack
coalescence, as in a
tensile test, produces a small value for $\beta$.
Such failures take place at weak spots with high local
stresses.  Any microscopic breaking stresses (fibers, bonds) will
have bounded distributions (e.g. Gaussian), so the
power-law distribution of energy implies that
the number of elements (i.e. size) of rupture
events has to vary in power-law manner.  In the peel-in nip test
(and also with tensile samples with a large notch)
the process reflects a more localized failure, but still scales as
a power-law albeit with a much higher exponent. 
In the peeling experiment, most likely the dynamics
is one-dimensional - the essential variation in the area covered by
events, corresponding to the energy takes place along the crack 
line but not into the sheet, in the planar direction. 

\begin{figure}\includegraphics[width=7cm
]{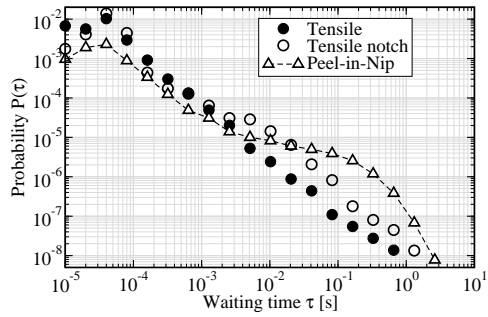}
\caption{Event interarrival time distributions for the
same cases as in the previous figure.} \label{fignelja}
\end{figure}

The variation in the exponents is analogous to the work by
Lockner et al., on fracture and crack growth in
granite \cite{Lockner}. They observed 
energy scalings with $\beta = 1.2 -- 2.3$, due to a setup
that allowed to follow the variation of $\beta$ during the fracture. 
In that context it
was suggested already earlier, by a model, that
$\beta$ should drop at maximum stress \cite{week}.
In addition the exponent was expected to recover after
the start of crack propagation, as was seen qualitatively
in the experiments of Lockner et al:
$\beta$ dropped from a value above 2 to near 1.2 at maximum
stress, and recovered back to a value of 1.7.
The waiting time distribution in tensile tests often results in a distinct
power-law \cite{oma}. With a large notch there is some
scattering from pure scaling behavior. In both cases the exponent
$\alpha$ is close to 1.0 (figure \ref{fignelja}). The peel-in-nip 
experiment diverges clearly from such an ideal dependence in the
time slot 10 - 500 ms. The fracturing takes place in a steady-state,
but such that maximum waiting times take into account that
the crack line has minimum velocity, imposed by the angular
velocity of the rolls.
The $\alpha$ values reported in literature are without exception in the
proximity of unity. For instance,
in a creep experiment with
a cellular glass material Omori's and Gutenberg-Richter laws were
observed, with the values for
$\alpha=1.3$ and $\beta=1.5$ \cite{Maas}. 
In AE experiments with ice $\alpha=1.0
\pm 0.3$ and $\beta = 1.3$ \cite{geofyysikoita}. These should be different
from the peeling
set-up, where the events by force take place at or very close to the
crack line.
Given $\alpha \sim 1$ the mechanism producing the scaling in all
these might presumably be universal.  



In order to study the origin of the typical scale
 in waiting time distribution,
we did peel-in-nip experiments with various strain rates. We observed
that the position of the ``leap length'' plateau shifts 
such that the lengthscale, interpreted as the distance
that the fracture line advances in a time corresponding to
the position of the plateau, stays roughly constant. It
is located between 1 $\mu$m and 50 $\mu$m, corresponding to a typical
fiber width and fiber-to-fiber bond scale in paper.
For a strain rate (500mm/min) the leap plateau
disappears. This  may originate from strain at the wire pulling
the rolls or inertia of the rolls, or the elastic energy stored into
strained part of the sample.
Without scaling the initial and final 
parts of the distributions overlap, which
implies the presence of other effects related to the fast
and slow timescales.

It is interesting to note that the Gutenberg-Richter exponents are close
in tensile tests with notch and peel-in-nip tests and 
that the Omori exponent
has close value in tensile and tensile with notch tests.
These experimental findings underline that these two power-laws are 
produced by
separate mechanisms. In particular it is possible to distort the
energy scaling, only.

\begin{figure}\includegraphics[width=7cm
]{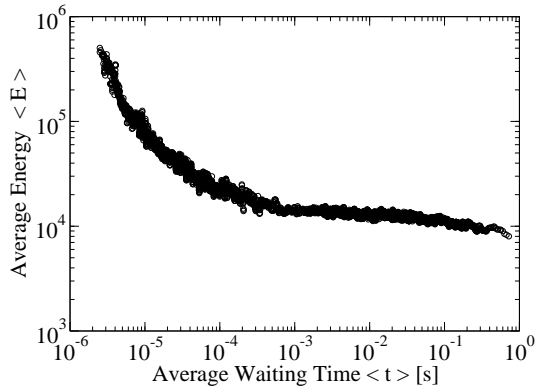}
\caption{Average energy (arbitrary units)
in a window of twenty consecutive
events, vs. the average waiting time in the same, for the
peel-in-nip test.} \label{figseit}
\end{figure}

\begin{figure}\includegraphics[width=7cm
]{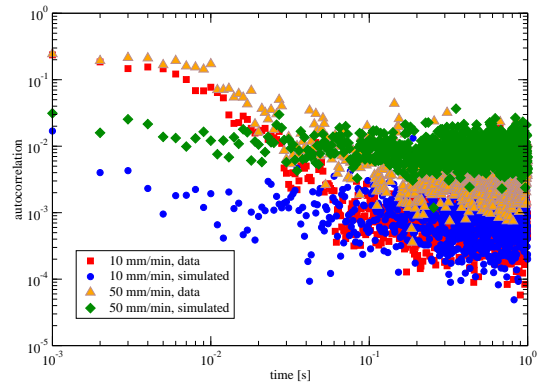}
\caption{Autocorrelations of the AE signal for two strain
rates. As a comparison, a randomize signal is presented for
both cases (see text).} \label{acf}
\end{figure}

For short enough time-windows the fracture
line may advance in a correlated fashion. This is demonstrated in
Fig.~\ref{figseit}. One observes that large events, in terms
of energy, are inter-correlated and more likely to follow each
other than if the signal was completely random. The natural
interpretation is that stronger crackline pinning is overcome
in a {\em correlated} fashion, so that the consecutive events
bear signatures of the energy stored due to the constant load
rate, during periods of no acoustic activity. One can also study
the autocorrelation functions (since the signal is stationary).
This results in a similar picture, in that the signal is correlated
upto a timespan which reaches about 0.05 s (depending on the straining
rate), as depicted in Fig.~\ref{acf}. Note that for the slower
rate the correlations extend further. The reference cases are
artificial signals using a randomized sequence of (exactly the
same) events and intervals.
We also attempted to correlate the AE data with the force
signal of the tensile testing machine; unfortunately it
became apparent that the effective force at the crackline
is masked by the two separate sheethalves that transmit
it from the rolls. In particular no direct correlated
between $E(t)$ and $F(t)$ could be detected.
The timeseries as such exhibited $1/f^\alpha$ noise, with
an exponent close but below unity. 

To conclude we have made experimental observations of acoustic
emission in paper peeling. The set-up is such that steady-state
crack propagation across a quasi-two dimensional material can
be followed. The main discoveries are a power-law for the
energies of the acoustic events, in spite of the fact that
the process is confined to an almost one-dimensional geometry,
and  for the event waiting time statistics. The former supports some
earlier observations, together with the tensile tests, that
a strong localization of the fracture process produces scalefree
statistics with a large exponent. The latter shows an
intriguing deviation from a power-law, which seemingly can be
attributed to the microstructure of paper.



\begin{figure}
\caption{Photograph of the peel-in-nip device, roll diameter is 80mm.} \label{figyksi}
\end{figure}

\begin{thebibliography}{99}
\bibitem{sethrev} 
J.P. Sethna, K.A. Dahmen, and C.R. Myers, 
	Nature (London) {\bf 410}, 242 (2001).
\bibitem{Bourev} Scalefree cracks are discussed
in: E.\ Bouchaud, J.\ Phys.\ Cond.\ Mat.\ {\bf 9}, 4319 
(1997).
\bibitem{Ciliberto}
A. Guarino, A. Garcimartin, and S. Ciliberto,
Eur. Phys. J. B {\bf 6}, 13 (1998);
A. Garcimartin {\em et al.}, 
Phys. Rev. Lett {\bf 79}, 3202 (1997).
\bibitem{penn}
L.C. Krysac and J.D. Maynard, 
Phys. Rev. Lett. {\bf 81}, 4428 (1998).
\bibitem{Petri2}
A. Petri {\em et al.}, 
	Phys. Rev. Lett {\bf 73}, 3423 (1994).
\bibitem{Lockner}
D.A. Lockner {\em et al.}, 
	Nature {\bf 350}, 39 (1991).
\bibitem{oma} L.I. Salminen, A.I. Tolvanen, and M.J. Alava,
Phys. Rev. Lett. {\bf 89}, 185503 (2002).
\bibitem{simu}
J.B. Rundle, S. Gross, W. Klein, C. Ferguson, D.L. Turcotte, 
Tectonophysics {\bf 277}, 147 (1997).
\bibitem{lopinaa}
J. Fineberg and M. Marder, Physics Reports {\bf 313}, 1-108 (1999).
\bibitem{line} e.g. T.\ Halpin-Healy and Y.-C.\ Zhang, Phys. Rep. {\bf 254},
215 (1995).
\bibitem{Fisher} S. Ramanathan, D. Erta\'s, and D.S. Fisher,
Phys. Rev. Lett. {\bf 79}, 873 (1997).
\bibitem{maloy}
J. Schmittbuhl and K.J. Målo$\!\!\!$/y, 
Phys. Rev. Lett. {\bf 78}, 3888 (1997); 
A. Delaplace, J. Schmittbuhl, and K. Målo$\!\!\!$/y, Phys. Rev. E {\bf 60},
 1337  (1999).
K.J. Målo$\!\!\!$/y, {\em et al.}, Int.
J. Fract. {\bf 121}, 9 (2003).

\bibitem{zapperi}
S. Zapperi, H.J. Herrmann, and S. Roux, Eur. Phys. J. B 
{\bf 17}, 131 (2000);
{\.A}str{\"o}m, M. Alava, and J. Timonen, Phys. Rev. E {\bf 62},  2878
 (2000); M.J. Alava and S. Zapperi,
Phys. Rev. Lett. {\bf 92}, 049601 (2004).
\bibitem{yama} 
T. Yamauchi, S. Okumura, N. Noguchi, Journal of Pulp and Paper
Science {\bf 16}, (1990).
\bibitem{Zapperi}P. K. Nukala, S. Simunovic, and S. Zapperi,
J. Stat. Mech. Theo. Expt., P08001 (2004).
\bibitem{recent} 
M. Minozzi, G.   Caldarelli, L. Pietronero, and
S. Zapperi, Eur. Phys. J. B {\bf  36}, 203 (2003).
\bibitem{herrmann} F. Tzschichholz and H.J. Herrmann,
Phys. Rev. {\bf E51}, 1961 (1995).
\bibitem{corte} 
H. Corte, {\em et al.}, in ``The Role of Fundamental
Research in Paper Making,'', Ed. J Brander, Mech. Eng.
Publ. Ltd., (London, 1983)), pp 571-584.
\bibitem{lys}
M.V. Lysak, Eng. Fract. Mech. {\bf 55}
443, (1996).
\bibitem{week} 
J.D. Weeks, D.A. Lockner, and J.D. Byerlee, Bull. Seism. Soc. Am.
{\bf 68}, 333-341 (1978).
\bibitem{Maas} 
C. Maes, {\em et al.}, Phys. Rev. B {\bf 57}, (1998).
\bibitem{geofyysikoita}
J. Weiss, J.R. Grasso,  and P. Martin, 
Proc. 6th Int. Conf. on AE/MS in Geol. Struct. \&
Mat., 1996, 583-595, Trans Tech Publications,
(Clausthal-Zellerfeld).

\end{thebibliography}
\end{document}